\documentclass[11pt,a4paper,reprint,notitlepage,superscriptaddress,aps,prl]{revtex4-2}

\usepackage{amsmath,amssymb,amsfonts}

\usepackage{amsthm}

\usepackage{bbold}

\usepackage{physics}

\usepackage[dvipsnames]{xcolor}
\usepackage{graphicx}

\usepackage{hyperref}
\hypersetup{
    colorlinks,
    citecolor=red,
    filecolor=red,
    linkcolor=red,
    urlcolor=red
}

\newcommand{\ba}{\begin{eqnarray}}
\newcommand{\ea}{\end{eqnarray}}
\newcommand{\lb}{\left(}
\newcommand{\rb}{\right)}
\newcommand{\be}{\begin{equation}}
\newcommand{\ee}{\end{equation}}

\usepackage[capitalise]{cleveref}

\begin{document}

\title{Exact Diagonalization of Sums of Hamiltonians and Products of Unitaries}


\author{Barbara \v{S}oda}
\email{Corresponding author: bsoda@perimeterinstitute.ca}
\affiliation{Department of Physics, University of Waterloo, Waterloo, ON N2L 3G1, Canada}
\affiliation{Perimeter Institute for Theoretical Physics,
Waterloo, ON N2L 2Y5, Canada}

\author{Achim Kempf}
\affiliation{Department of Physics, University of Waterloo, Waterloo, ON N2L 3G1, Canada}
\affiliation{Perimeter Institute for Theoretical Physics,
Waterloo, ON N2L 2Y5, Canada}
\affiliation{Department of Applied Mathematics, University of Waterloo, Waterloo, ON N2L 3G1, Canada}
\affiliation{Institute for Quantum Computing, University of Waterloo, Waterloo, ON N2L 3G1, Canada}

\begin{abstract}
We present broadly applicable tools for tracking the behavior of eigenvalues and eigenvectors under the addition of self-adjoint operators and under the multiplication of unitaries, in finite-dimensional Hilbert spaces.  
The new tools provide explicit non-perturbative expressions for the eigenvalues and eigenvectors. 
To illustrate the broad applicability of the new tools, we outline several applications, for example, to Shannon sampling in information theory. A longer companion paper applies the new tools to adiabatic quantum evolution, thereby shedding new light on the connection between an adiabatic quantum computation's usage of the resource of entanglement and the quantum computation's speed.

\end{abstract}

\maketitle
The behavior of eigenvalues and eigenvectors plays important roles throughout science and engineering \cite{LA0,LA1,LA2}. Of particular interest is their behavior under the addition of Hermitian operators, such as Hamiltonians, and under the multiplication of unitaries, such as time evolution operators. 
In the literature, these problems are often investigated perturbatively since nonperturbative results, such as level repulsion \cite{LR0,LR1,LR2} and Cauchy interlacing \cite{CI0,CI2}, 
are few in number \cite{Denton_2021}. 

Here, we present new non-perturbative tools that yield the exact behavior of the eigenvalues and eigenvectors of Hermitian operators under addition, and of unitary operators under multiplication, in finite-dimensional Hilbert spaces. 
In order to make our results as accessible as possible, we include here only a sketch of the proofs. The underlying calculations and the proofs are presented in the Supplemental Material. 

The new results provide tools that are broadly-applicable and, as an example application, we use them to generalize Shannon sampling in information theory. Further, in the companion paper \cite{gabbassov2024}, the new tools are applied to adiabatic quantum evolution, thereby shedding new light on the connection between the speed of an adiabatic quantum computation and its usage of the resource of entanglement. 

\bf The addition of Hermitian operators. \rm We begin by considering an arbitrary Hermitian operator, $S$, with a generic, i.e., nondegenerate spectrum, acting on an $N$-dimensional Hilbert space $\cal{H}$. (We denote the operator by the letter $S$ rather than $H$ because it does not need to be a Hamiltonian). We ask how the eigenvalues and eigenvectors of $S$ change when adding to it an arbitrary Hermitian operator $R$. To this end, by the spectral theorem, we decompose $R$ into a weighted sum of rank 1 projectors: $R=\sum_{j=1}^N \mu^{(j)} \vert v^{(j)}\rangle\langle v^{(j)}\vert$. 
Since the sum $S+R$ can be obtained by successively adding these weighted projectors to $S$, we first focus on 
how the spectrum of an operator, $S$, changes under the addition of a single weighted projector: 
\be S(\mu):=S+\mu \vert v\rangle\langle v\vert
\label{e1}
\ee
For examples of prior results on \cref{e1}, see \cite{stats1,stats2} for the special case of $S$ being a random matrix, and see \cite{sorensen,gu1994stable,huang2010nonlinear} for the special case of real vector spaces and for numerical methods. For a review, see, e.g., \cite{simon}.
Here, we derive new nonperturbative results by developing a unifying theory of both the addition of Hermitian operators  
and the multiplication of unitaries, and by applying the Lagrange inversion theorem.

To this end, we let $\vert v\rangle$ be an arbitrary fixed normalized vector in $\cal{H}$ and we let the weight $\mu$ range from $-\infty$ to $+\infty$. 
We denote the sets of eigenvalues and eigenvectors of $S$ by $\{s_n\},\{\vert s_n\rangle\}$ with the ordering $s_1<s_2<\ldots <s_N$. 
Correspondingly, we denote the sets of eigenvalues and eigenvectors of $S(\mu)$ by $\{s_n(\mu)\},\{\vert s_n(\mu)\rangle\}$, i.e., we have $s_n(0)=s_n,\vert s_n(0)\rangle=\vert s_n\rangle$. 

\bf Trivial special cases. \rm If $\vert v\rangle$ is chosen so that  $\langle s_m\vert v\rangle = 0$ for some $m$, then the eigenspaces of $S$ spanned by these vectors  $\vert s_m\rangle$ remain unaffected, i.e., their eigenvalues and eigenvectors do not change with $\mu$: $s_m(\mu)=s_m, \vert s_m(\mu)\rangle =\vert s_m\rangle ~\forall \mu\in\mathbb{R}$. If we choose $\vert v\rangle$ to be an eigenvector, $\vert v\rangle:=\vert s_r\rangle$, then
$\langle s_m\vert v\rangle = 0~~ \forall m\neq r$ and all eigenvectors and eigenvalues are frozen, except that  $s_r(\mu)=s_r+\mu, \vert s_r(\mu)\rangle =\vert s_r\rangle ~\forall \mu\in\mathbb{R}$.
\newline \bf Qualitative behavior of the eigenvalues. \rm 
We now consider the generic case in which $\vert v\rangle$ is not orthogonal to any eigenvectors of $S$. 
As we show in the Supplement, and as is illustrated in \cref{fig1}, in this case, as we let $\mu$ (on the vertical axis) increase from $\mu=-\infty$ to $\mu=+\infty$, all the eigenvalues $s_n(\mu)$ of $S(\mu)$ strictly monotonously increase, i.e., they move left to right on the $s$-axis, bumping each other forward in the process. In fact, their behavior is akin to that of the balls of a Newton's cradle: The leftmost `ball' $s_1(\mu)$ starts from $s=-\infty$ (for $\mu=-\infty$), then increase to reach  $s(\mu)=s_1$ at $\mu=0$, after which it further increases and converges to a value $s_1^*$ (as $\mu\rightarrow +\infty$). As, $s_1(\mu)$ approaches $s_1^*$, it almost touches the second eigenvalue (or `ball'), $s_2(\mu)$, but instead repels $s_2(\mu)$ and pushes it to the right. Similarly, as the second ball $s_2(\mu)$ increases and converges toward a value $s_2^*$, it pushes the third ball, $s_3(\mu)$, and so on. Finally, the last eigenvalue, $s_N(\mu)$ is sent off to infinity as $\mu\rightarrow \infty$. Hence, if we define $s_0^*:=-\infty$ and $s_N^*:=+\infty$, we have that each 
eigenvalue $s_n(\mu)$ for $1\le n\le N$ starts from a value $s^*_{n-1}$ (for $\mu=-\infty$), moves through $s_n$ (as $\mu=0$) and then converges to a value $s_n^*$ (for $\mu\rightarrow +\infty$). 
\newline
\bf Exact behavior of the eigenvalues. \rm 
Due to the Newton cradle-like behavior of the eigenvalues, the full set of eigenvalues $\{s_n(\mu)\}$ for all $n=1\ldots N$ and all $\mu\in\mathbb{R}\cup \{+\infty\}$ cover the real line exactly once. For any $s$, there therefore exists a unique pair $(n,\mu)$ such that $\vert s\rangle:=\vert s_n(\mu)\rangle$ obeys $S(\mu)\vert s\rangle = s\vert s\rangle$.  
Defining $v_n:=\langle s_n\vert v\rangle$, for any $s\in\mathbb{R}$, the corresponding $\mu$ reads:
\begin{equation}
  \mu(s) = \left( \sum_{n=1}^N\frac{\abs{v_n}^2} {s-s_n}\right)^{-1} \label{cauchy}
\end{equation}
Using \cref{cauchy} we obtain, for every choice of energy eigenvalue $s$, the corresponding value of the coupling constant $\mu$. It would be highly desirable to invert this relation to obtain for every value of the coupling constant $\mu$ the corresponding spectrum $\{s_n(\mu)\}$. This would appear to be impossible because Galois theory shows that, for $N>4$, the eigenvalues, being the zeros of the characteristic polynomial of $S(\mu)$, cannot be obtained as a finite combination of addition, subtraction, multiplication, division, and taking 
k-th roots. However, as we show below, the eigenvalues $s_n(\mu)$ can be obtained as a power series in $\mu$ with an infinite number of terms where, using the Lagrange inversion theorem, we can give the exact expression for each coefficient explicitly. 

\bf Role of the $s_n^*$. \rm  \cref{cauchy} implies that the values $s_n^*$ for $n=1\ldots (N-1)$ are the solutions to: 
\begin{equation}
  \sum_{m=1}^N\frac{\abs{v_m}^2} {s_n^*-s_m}=0 \label{sstar}
\end{equation}
As we show in the Supplement, the $s_n^*$ can therefore be identified as the interlaced eigenvalues whose mere existence is implied by Cauchy's interlacing theorem. In this context, we also show in the Supplement that the well-known \it Eigenvectors from Eigenvalues \rm result of Denton, Parke, Tao and Zhang, \cite{Denton_2021}, follows straightforwardly. 
\begin{figure}[t!]
	\centering
	\includegraphics[width=0.81\columnwidth]{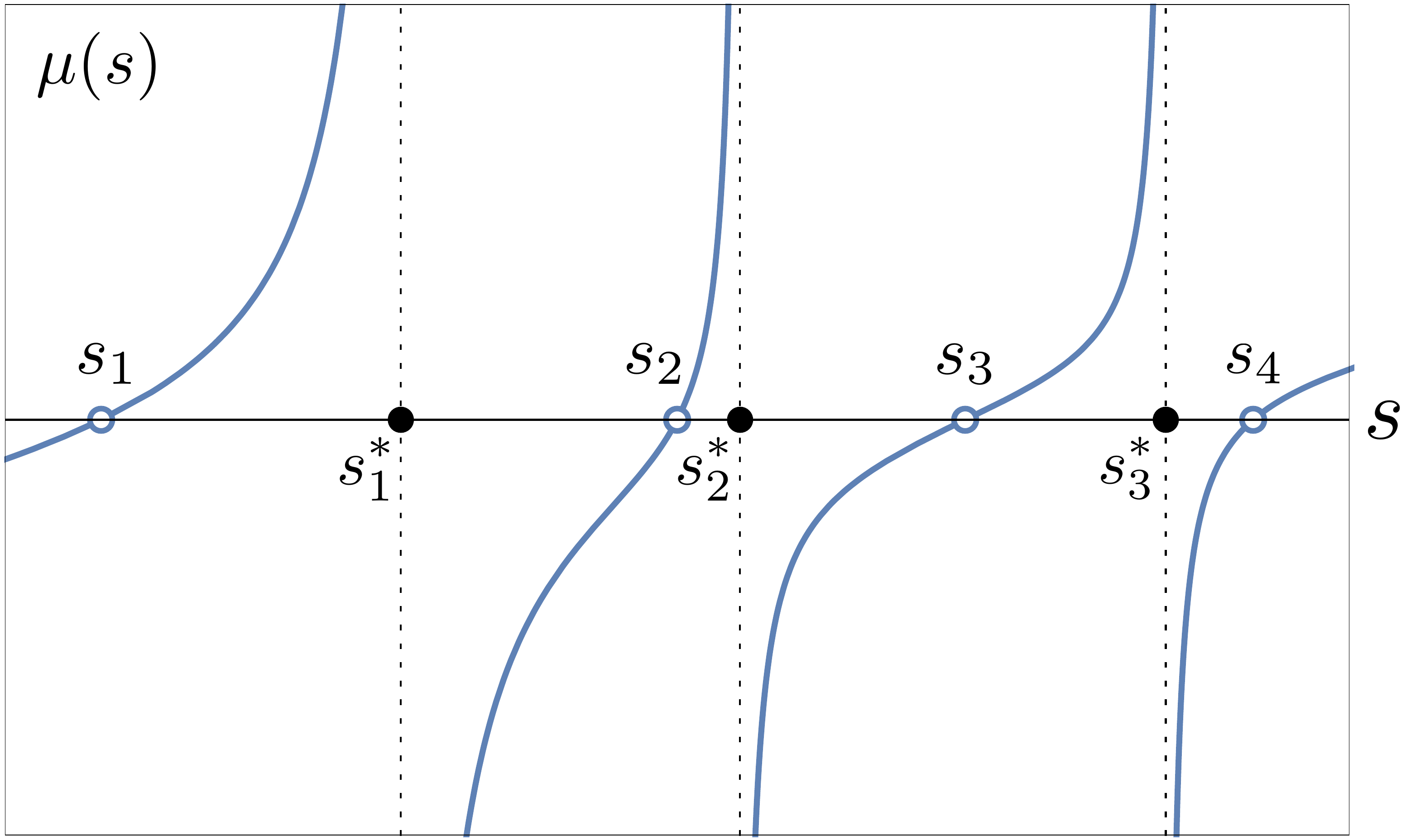}
	\caption{\label{fig1}
		Plot of $\mu(s)$, using \cref{cauchy}, for a generic example with $N=4$. From bottom to top, the curves show the evolution of the four eigenvalues $\{s_n(\mu)\}_{n=1}^4$ with increasing $\mu$. For $\mu=0$, the eigenvalues are $s_1$, $s_2$, $s_3$, $s_4$. As $\mu\rightarrow +\infty$, the eigenvalues tend to $s_1^*$, $s_2^*$, $s_3^*$ and $s_4^*=+\infty$ respectively. }
\end{figure}

\bf Velocity of the eigenvalues. \rm     
Differentiating \cref{cauchy} yields:
\be
\frac{d\mu(s)}{ds}=\left( \sum_{m=1}^N\frac{\abs{v_m}^2} {s-s_m}\right)^{-2} \sum_{r=1}^N \frac{\abs{v_r}^2}{\lb s-s_r\rb^2} \label{npvel}
\ee
The inverse, $ds/d\mu$, is the velocity of the eigenvalues, $s$, with respect to $\mu$. Thus, the special case of the velocity $ds_n(\mu)/d\mu$ of an eigenvalue $s_n(\mu)$ at $\mu=0$ reads 
\be
\frac{\dd s_n\lb \mu\rb}{\dd \mu}\eval_{\mu = 0}=\abs{v_n}^2  \label{vel2}
\ee
and hence $ds_n(\mu)/d\mu = |\langle s_n(\mu)\vert v\rangle|^2$ for all $\mu$. Therefore, the velocities stay positive for all $\mu$: since $ds_n/d\mu=0$ would imply $\langle s_n(\mu)\vert v\rangle =0$, then $s_n(\mu)$ would be frozen for all $\mu$, in contradiction to the assumption that $\langle s_n\vert v\rangle \neq 0$. 
Since $\sum_{m=1}^N\vert v_m\vert^2=1$, the velocities always sum to one. Indeed, since $d\Tr(S(\mu))/d\mu =\Tr(\vert v\rangle\langle v\vert)=1$ for all $\mu$, the eigenvalues conserve their total momentum, as in a Newton cradle, in the sense that for all $\mu\in\mathbb{R}:~~\sum_{n=1}^N\frac{\dd s_n\lb \mu\rb}{\dd \mu}=1$. 
\newline \bf Behavior of spectra under the multiplication of unitaries. \rm
An analogous result holds for unitaries. Let us consider an arbitrary fixed unitary $U$ acting on $\cal{H}$. Instead of adding a $1$-parameter family of  projectors as we did above, we now multiplicatively act on $U$ from the left with a $U(1)$ group. We let the elements of this $U(1)$ act as the identity everywhere except on the dimension spanned by some arbitrary fixed vector, $\vert  w\rangle$, i.e., we multiply $U$ from the left with the $U(1)$-group of unitaries $(\mathbb{1}+\left(e^{i \alpha}-1\right) \ket{w}\bra{w})$ for $\alpha \in [0,2\pi)$ to obtain a family of unitaries $U(\alpha)$:
\begin{equation}
    U({\alpha}):=(\mathbb{1}+\left(e^{i \alpha}-1\right) \ket{w}\bra{w})\  U \label{unite}
\end{equation}
Let us denote the eigenvalues of $U$ by $\{u_n\}$, ordered counterclockwise on the complex unit circle, starting from $1$. We denote the eigenvalues of $U(\alpha)$ by $u_n(\alpha)$, i.e., we have $u_n(0)=u_n$. We define 
$w_n:=\langle u_n\vert w\rangle$.
Clearly, if $\vert w\rangle$ is chosen to obey $\langle u_m\vert w\rangle=0$ for some $m$ then the corresponding eigenvalues and eigenvectors are frozen, $u_m(\alpha)=u_m, ~\vert u_m(\alpha)\rangle =\vert u_m\rangle~\forall \alpha$. Excluding these trivial cases, we can show again a Newton cradle like behavior: as $\alpha$ runs from $0$ to $2\pi$, each eigenvalue $u_n(\alpha)$ runs counterclockwise on the complex unit circle, reaching $u_{n+1}$ as $\alpha\rightarrow 2\pi$. Except, $u_N(\alpha)$ runs towards $u_1(\alpha)$.

\bf Equivalence of the unitary and Hermitian cases. \rm We can show that the left multiplication of a unitary operator by a representation of the group $U(1)$ according to \cref{unite} is equivalent to the addition of a weighted rank $1$ projector to a Hermitian operator. The equivalence is established by this Cayley transform:
\begin{equation}
    U(\alpha) = (S(\mu) - i\mathbb{1})(S(\mu)+i\mathbb{1})^{-1} \label{Cayley1}
\end{equation}
\begin{equation}
    S(\mu) = - i (U(\alpha) +\mathbb{1})(U(\alpha) - \mathbb{1})^{-1} \label{Cayley2}
\end{equation}
Let us recall here that the addition of Hermitian operators does generally not amount to the multiplication of the corresponding unitaries when mapping Hermitian operators to unitary operators by exponentiation. Instead, one then needs to deal with the unwieldy Baker Campbell Hausdorff formula \cite{BCH}. In contrast, we have shown here that by mapping Hermitian operators to unitaries via the Cayley transform, the addition of Hermitian operators (projector by projector) to a given Hermitian operator exactly amounts to the left multiplication of unitaries ($U(1)$ element by $U(1)$ element) onto a corresponding unitary.  

Further then, from \cref{Cayley1,Cayley2}, $U(\alpha)$ and $S(\mu)$ possess the same eigenspaces, and their eigenvalues are related by a M\"obius transform: $u_n(\alpha)=(s_n(\mu)-i)/(s_n(\mu)+i)$.
We can show that
\begin{equation} 
\ket{v}=\frac{2}{\sqrt{\sum_{m=1}^N \abs{w_m}^2 \lb s_m^2+1\rb}}\lb \mathbb{1}- U\rb ^{-1} \ket{w}
\end{equation}
and in coefficients:
\begin{eqnarray}
v_n&=&\frac{1}{\sqrt{\sum_{m=1}^N \abs{w_m}^2 \lb s_m^2+1\rb}} \frac{s_n+i}{i} w_n\\
w_m&=&\frac{1}{\sqrt{\sum_k \frac{\abs{v_k}^2}{s_k^2+1}}}\frac{i}{s_m+i}v_m
\end{eqnarray}
Further, $\mu$ and $\alpha$ are related by:
\begin{eqnarray}
\alpha(\mu) &=& 
2 \arccot\left[\lb\sum_k\frac{\abs{v_k}^2}{s_k^2+1}\rb^{-1}\lb\frac{1}{\mu}+\sum_m\frac{\abs{v_m}^2 s_m}{s_m^2+1}\rb\right] \nonumber
\\
\mu\lb \alpha \rb &=&
\lb  \sum_{m=1}^N \frac{\abs{v_m}^2}{ s_m^2+1}  \cot \lb \frac{\alpha}{2}\rb -\sum_{k=1}^N \frac{\abs{v_k}^2 s_k}{ s_k^2+1}\rb^{-1}
\end{eqnarray}
We read off that $\mu(\alpha)=0$ for $\alpha=0$ and that, as $\alpha$ increases
from $0$ to the finite value $\alpha^*:=2\arccot(\sum_k\vert w_k\vert^2s_k)$, we have that $\mu(\alpha)$ runs to $+\infty$. As $\alpha$ further increases, $\mu(\alpha)$ comes back up from $-\infty$ and finally $\mu(\alpha)\rightarrow 0$ as $\alpha\rightarrow 2\pi$. 


\bf Exact behavior of the eigenbases. \rm 
The normalized eigenvectors $\vert s\rangle$ are defined only up to phases and so is their inner product. But the absolute value of their overlap is unique and we can prove that for all $s\in \mathbb{R}$: 
\begin{equation} \label{evec}
\abs{\braket{s}{s_n}}=\frac{\abs{v_n}}{\abs{s-s_n}} \lb\sum_{m=1}^N \frac{\abs{v_m}^2} {(s-s_m)^2}\rb^{-1/2}
\end{equation}
It is possible to choose the phases of the eigenvectors such that the overlap function $\langle s\vert s_n\rangle$ is real and continuous:
\begin{equation}\label{evecs}
\braket{s}{s_n}=\frac{(-1)^n \abs{v_n}}{s-s_n} \lb\sum_{m=1}^N \frac{\abs{v_m}^2} {(s-s_m)^2}\rb^{-1/2} \prod_{r=1}^N \lb -1\rb^{\theta\left( s
-s_r \right)}
\end{equation}
Here, the singularity at $s=s_n$ is trivially removable and $\theta$ is the Heaviside function with $\theta(0):=0$.
Notice that, from \cref{evecs}, the  eigenvectors of $S(\mu)=S+\mu \vert v\rangle\langle v\vert$ take the  form $\vert s\rangle\propto \sum_{n=1}^N(-1)^n\vert v_n\vert(s-s_n)^{-1}\vert s_n\rangle$.
\newline
\bf Iteration to obtain $\bf{S+R}$. \rm
The Newton cradle for $\vert v\rangle := \vert v^{(1)}\rangle$ and $\mu:=\mu^{(1)}$ yields the first step in the successive addition of the weighted projectors 
$\mu^{(j)} \vert v^{(j)}\rangle\langle v^{(j)}\vert$
to $S$ to obtain $S+R$. 
For the second step, we need the eigenvalues 
of the operator $S+\mu^{(1)}\vert v^{(1)}\rangle\langle v^{(1)}\vert$.
In the Supplement, we use the Lagrange inversion theorem, see, e.g., \cite{jackson}, to calculate these eigenvalues nonperturbatively. The result is an expression for the eigenvalues that is a power series with all Taylor coefficients explicitly given. (Recall that by Galois theory, the power series must contain an infinite number of nonzero coefficients). Our result (which, as all our results, we also tested numerically) reads:
\begin{eqnarray}
s_r(\mu) &=& s_r \nonumber \\ & &-\sum_{n=1}^\infty \frac{\mu^n}{n}\!\!\!\!\!\!
\sum_{\substack{k_1,...,k_n=0\\ \sum_{i=1}^n k_i=n-1}}^{n-1}\sum_{\substack{p_1,...,p_n=1\\ p_1\neq r,...,p_n\neq r}}^N\prod_{i=1}^n\frac{\vert v_{p_i}\vert^2-\frac{\delta_{k_i,0}}{N-1}}{(s_r-s_{p_i})^{k_i}} \nonumber
 \end{eqnarray}
In contrast to conventional Rayleigh-Schrödinger perturbation theory, this expression is non-perturbative in the sense that it is closed-form and non-recursive, i.e., to write down the $n$'th Taylor coefficient, the prior coefficients are not needed. The Lagrange inversion theorem guarantees that the radius of convergence of the power series is larger than zero. If it is finite, multiple uses of the series may be needed to let $\mu$ run from $0$ to $\mu^{(1)}$. 

For the second step, i.e., to add the next weighted projector to $S$, we now need the
coefficients $v^{(2)}_n$ of $\vert v^{(2)}\rangle$ in the eigenbasis $\{\vert s^{(1)}_r(\mu^{(1)})\rangle\}_{n=1}^N$. Using \cref{evecs}, we obtain them exactly: 
\be
v^{(2)}_n = \langle s_n(\mu^{(1)})\vert v^{(2)}\rangle =\sum_{r=1}^N\langle s_n(\mu^{(1)})\vert s_r\rangle\langle s_r\vert v^{(2)}\rangle \label{step2}
\ee
These exact methods now allow us to successively turn on all the projectors that comprise $R$ to obtain $S+R$ as the result of $N$ Newton cradles. 
\newline
\bf Sketch of the proofs. \rm The derivations and proofs are given in the Supplemental Material. They are inspired by \cite{AK-Shannon1,AK-Shannon2,AK-Shannon3,AK-Shannon4,AK-Hao}, by von Neumann's theory of self-adjoint extensions of simple symmetric operators \cite{vNeumann,Akhiezer}.
Here is a sketch of how the results are obtained. 
A priori, von Neumann's theory of self-adjoint extensions is only applicable in infinite-dimensional Hilbert spaces because there are no self-adjoint extensions in finite dimensions. But von Neumann's theory can be generalized to finite dimensions after formulating it as the theory of unitary extensions of isometric operators. We then find that, unexpectedly, this generalization describes for deficiency indices $(1,1)$, via the Cayley transform, a family of self-adjoint operators of the form $S(\mu) =S+\mu R$ with $R$ being of rank 1. This allows us to calculate the spectra and eigenvectors, where $\mu$ can be calculated nonperturbatively as a function of the eigenvalues. The goal is of course to calculate the converse, i.e., the eigenvalues as a function, $s_n(\mu)$, of $\mu$ and, as discussed above, this can be done explicitly by using the Lagrange inversion theorem.  
\newline
\bf Special case: Level repulsion. \rm 
The important phenomenon of level repulsion, see, e.g., \cite{von1993merkwurdige,LR3,LR4}, arises as a special case: Let us consider a $\vert v\rangle$ which obeys $v_m=\langle s_m\vert v\rangle = 0$ for one $m$. Then the corresponding eigenvalue is frozen, $s_m(\mu)=s_m~\forall \mu$. The remaining eigenvalues $s_n(\mu)$ for $n\neq m$ form a Newton cradle and $s_{m-1}(\mu)$ will cross $s_m$ as $\mu$ runs from $0$ to $\infty$. In general, however, $\vert v_m\vert\neq 0$. In this case, no matter how small $\vert v_m\vert$ is, the eigenvalue $s_m(\mu)$ does participate in Newton's cradle and is, therefore, not being crossed as $\mu$ runs. For very small $\vert v_m\vert$, the eigenvalue $s_{m-1}(\mu)$ can at most closely approach $s_m(\mu)$ while $s_m(\mu)$ barely moves until eventually $s_{m-1}(\mu)$ must repel $s_m(\mu)$ to send it on its way to $s_{m}^*$ as $\mu\rightarrow\infty$. 
Let us now consider the generating of the sum of two Hermitian operators, $S+R$, by successively continuously turning on one after the other of the weighted projectors that comprise $R$. 
Since each turning on of a weighted projector is a Newton cradle process,
we conclude that the eigenvalues in generic cases cannot cross during that process, i.e., we must have level repulsion, except when and only when the  projector $\vert v^{(i)}\rangle\langle v^{(i)}\vert$ that is being added is exactly orthogonal to an eigenvector of the operator that it is being added to.
\newline
\bf Example application: Generalization of Shannon sampling. \rm 
Shannon sampling theory is central in information theory, where it establishes the equivalence of continuous and discrete representations of information \cite{Shannon,Shannon0,Shannon1,Shannon2,Shannon3}. Applied to physics, Shannon sampling shows that in the presence of a suitable natural UV cutoff, spacetime could be simultaneously discrete and continuous in the same way that information can, \cite{AK-Shannon2,AK-spacetime2,AK-spacetime3}. Shannon sampling also possesses a close relationship to generalized uncertainty principles \cite{AK-GUP1,AK-Shannon1}. We now show that the present results yield a generalization of Shannon sampling that is advantageous in that it accounts for varying information densities (a form of varying bandwidth) while also eliminating truncation errors for finite-length signals.

The basic Shannon sampling theorem \cite{Shannon}
concerns $\Omega$-bandlimited functions, i.e., functions $f$ for which there exists an $\tilde{f}$ so that $f(s)=\int_{-\Omega}^\Omega e^{i\omega s} \tilde{f}(\omega)d\omega$. The theorem states that if the amplitudes $\{f(s_n)\}_{n=-\infty}^\infty$ of an $\Omega$-bandlimited function $f$ on the real line are sampled at the so-called Nyquist spacing $\pi/\Omega$, e.g., $s_n:=n\pi/\Omega~~\forall n\in \mathbb{Z}$, then $f$ can be exactly reconstructed  from these samples for all $s\in\mathbb{R}$:
\be
f(s) =\sum_{n=-\infty}^\infty \label{shannon} f(s_n)~\frac{\sin(s\Omega -n\pi)}{s\Omega -n\pi}
\ee
While abundantly useful, this theorem has the drawbacks of assuming a constant bandwidth and a correspondingly constant Nyquist rate as well as requiring an infinite number of samples. In practice, these drawbacks can lead to inefficiencies and truncation errors respectively. In the literature, Shannon sampling has been generalized to varying Nyquist sampling rates, corresponding to varying bandwidths, see, e.g., \cite{AK-Shannon1,AK-Hao,AK-GUP2}. However, these results still require the taking of infinitely many samples, the obstacle being the use of von Neumann's method of self-adjoint extensions, which requires infinite dimensions. 

Here, the present results  generalize von Neumann's theory to finite dimensions and we now show that they can be applied to generalize Shannon sampling theory to enable both variable Nyquist rates and finite numbers of samples without incurring truncation errors. To this end, let us assume given a Newton cradle, i.e., a family of Hermitian operators $S(\mu) = S+\mu \vert v\rangle\langle v\vert$ for fixed $S$, fixed $\vert v\rangle$ and for $\mu$ running through $\mathbb{R}$. We can then uniquely associate to every vector $\vert f\rangle$ in the Hilbert space the function $f(s):=\langle s\vert f\rangle$. These functions obey
\be
f(s) = \langle s\vert f\rangle =\sum_{n=1}^N\langle s\vert s_n\rangle\langle s_n\vert f\rangle=\sum_{n=1}^N\langle s\vert s_n\rangle~f(s_n) \label{newsampling}
\ee
and more generally, for any $\mu$:
\be
f(s) = \sum_{n=1}^N\langle s\vert s_n(\mu)\rangle~f(s_n(\mu)) \label{newsampling2}
\ee
Here, $\langle s\vert s_n(\mu)\rangle$ can be assumed real and continuous as given in \cref{evecs}. \cref{newsampling} thereby beautifully generalizes and de-mystifies sampling theory: The reason why a bandlimited continuous function can be perfectly reconstructed from discrete samples is that knowing the coefficients (i.e., here the sample values) of a vector, $\vert f\rangle$, in the eigenbasis of one Hermitian operator $S(\mu)$ implies knowing the vector itself, and therefore implies being able to calculate its  coefficients in all bases, including in the eigenbases of all other $S(\mu')$, which yields the bandlimited function everywhere. Further, \cref{newsampling2} shows that the function $f(s)$ can be reconstructed everywhere, i.e., for all $s\in\mathbb{R}$, from the samples taken on the discrete spectrum of \it any \rm of the operators $S(\mu)$, thereby yielding the equivalence of the continuous  and the discrete representations of the information contained in $\vert f\rangle$. 
It is straightforward to see that this result generalizes the sampling theory of \cite{AK-Shannon1,AK-Hao,AK-GUP2} by allowing not only variable Nyquist rates but also finitely many samples without incurring truncation errors, while recovering the sampling results of \cite{AK-Shannon1,AK-Hao,AK-GUP2} in the limit $N\rightarrow\infty$. In turn, \cite{AK-Shannon1,AK-Hao,AK-GUP2} yield the Shannon sampling theorem \cite{Shannon} as the special case of an eternally constant Nyquist rate. 

It will be interesting to explore applications of the new sampling result given in \cref{evecs} and \cref{newsampling} in circumstances with known varying Nyquist rate, such as in synthetic aperture astronomy. For example, in the planned SKA experiment \cite{SKA1}, the effective bandwidth between any pair of antennas depends on their apparent distance as seen by the observed object and therefore varies predictably with the earth's rotation. Since large communication costs demand maximally efficient data taking, the generalized Shannon sampling method could be useful by enabling sampling and reconstruction at maximally efficient continuously-adjusted Nyquist rates, without incurring truncation errors at the beginning and end of sample taking.  
\newline

\bf Outlook. \rm  
We here point, in particular, to the follow-up companion paper \cite{gabbassov2024} in which, building on the present results, the dynamics of entanglement during adiabatic quantum evolution is nonperturbatively analyzed. Applied to adiabatic quantum computing, \cite{AQC1,AQC2} (which is  equivalent to algorithmic quantum computing \cite{Nielsen} up to polynomial overhead) \cite{AQC3,aharonov2008adiabatic}), the companion paper \cite{gabbassov2024} then shows that the dynamical changes in entanglement among subsystems can be traced to the avoided level crossings of the Newton cradle like behavior of eigenvalues: it is at these avoided level crossings that eigenvalues trade their eigenvectors, thereby in this sense `weaving' the entanglement among subsystems. This then sheds new light on the relationship between the narrowness of avoided level crossings, and therefore the speed of an adiabatic quantum computation, and the usage of that quantum computations resource of entanglement, which in turn directly depends on the ruggedness of the cost function landscape.   

Beyond adiabatic quantum computing, the present results could yield insights into effective forces \cite{Fulling,Birrell} that arise from ground state dynamics, such as Casimir forces \cite{Casimir0,Casimir1}, and gravity in Sakharov’s approach \cite{Sakharov1}. 
More generally, the new results on eigenvalue dynamics  could yield fresh insights into spectral geometry \cite{SGeometry0}, i.e., into the dynamics of spectra of wave operators as a function of parametrized changes to the metric, building on \cite{AK-Shannon3,SGeometry2,replacing}. Further, in quantum communication, the application of the new results to the dynamics of density matrices could yield new insights into the quantum channel capacities of interactions, building, e.g., on \cite{Qcapacity}. Also, when applied to random matrix theory, \cite{rmt0}, a Newton cradle type analysis could help shed light on the dynamics of the BBT transition \cite{rmt1,rmt2} in quantum chaotic systems. 

It will also be interesting to explore to what extent the present results can aid in numerical studies, see e.g., \cite{sorensen}, the generalization to higher deficiency indices, and 
the limit $N\rightarrow \infty$. Regarding the latter, what is clear so far is that if the Cayley transforms of the unitaries $U(\alpha)$ are unbounded Hermitian operators then by von Neumann's theory of self-adjoint extensions, the operators $S(\mu)$ differ no longer by a multiple of a projector but by a domain extension, such as a boundary condition in the case of differential operators \cite{Akhiezer}. Interestingly they can also differ by a Hermitian operator, namely when using an auxiliary Hilbert space \cite{posilicano}. 
\bigskip\newline 
\bf Acknowledgements. \rm 
The authors are grateful for valuable feedback on the manuscript from Jason Pye, Marcus Reitz, Koji Yamaguchi and Eduardo Mart\'in-Mart\'inez. AK acknowledges support through a grant from the National Research Council of Canada (NRC), a Discovery Grant of the National Science and Engineering Council of Canada (NSERC), a Discovery Project grant of the Australian Research Council (ARC) and a Google Faculty Research Award. 
AK and B\v{S} are supported in part by the Perimeter Institute, which is supported in part by the Government of Canada through the Department of Innovation, Science
and Economic Development Canada and by the Province of Ontario through the Ministry of Economic Development, Job Creation and Trade.

\bibliographystyle{apsrev4-2}
\bibliography{bibliography}
\newpage

\section{Supplemental Material}

We describe here the key elements of the derivations and proofs. They are inspired by \cite{AK-Shannon1,AK-Shannon2,AK-Shannon3,AK-Shannon4,AK-Hao}, by von Neumann's theory of self-adjoint extensions of simple symmetric operators \cite{vNeumann,Akhiezer}, and one proof uses the Lagrange inversion theorem. We also show that Cauchy's interlacing theorem arises as a special case, and we also show that the well-known \it Eigenvectors from Eigenvalues \rm result \cite{Denton_2021} follows straightforwardly. 

A key inspiration for our present results has been the fact that while von Neumann's theory of self-adjoint extensions is only applicable in infinite-dimensional Hilbert spaces, because there are no self-adjoint extensions in finite dimensions, von Neumann's theory can be generalized to finite dimensions after formulating it as the theory of unitary extensions of isometric operators. We then find that, unexpectedly, this generalization describes, via the Cayley transform, the diagonalization of self-adjoint operators of the form $S(\mu) =S+\mu R$, including the elementary case where $R$ is of rank 1. In this case, $\mu$ can be calculated nonperturbatively as a function of the eigenvalues. The goal is of course to calculate the converse, i.e., the eigenvalues as a function, $s_n(\mu)$, of $\mu$. Galois theory implies that $s_n(\mu)$ cannot be obtained as a finite expression in radicals in terms of $\mu$. However, we show that the solution, i.e., the eigenvalues $s_n(\mu)$ as a function of $\mu$, can be obtained exactly and explicitly as a power series by using the Lagrange inversion theorem. The generic case of $S(\mu) =S+\mu R$ without rank restrictions then follows.

We will, therefore, map self-adjoint to unitary operators, and vice versa, using the Cayley transform. An advantage of the Cayley transform is that, unlike exponentiation, the Cayley transform is bijective and hence uniquely invertible. Note that, since we work here in finite-dimensional Hilbert spaces, the terms Hermitian and self-adjoint can be used interchangeably.    
\section{Relation between Hermitian and unitary Newton cradles}
We will now prove that the left action of a representation of the unitary group $U(1)$ on a unitary operator $U$ is mapped, via the Cayley transform, into the addition of multiples of a rank $1$ projector to a Hermitian operator $S$. Concretely, assume $U$ is an arbitrary fixed unitary acting on a finite-dimensional Hilbert space $\cal{H}$. Then its Cayley transform is defined to be the Hermitian operator $S$:
\begin{equation}
    S:=-i \lb  U + \mathbb{1}\rb \lb U-\mathbb{1}\rb^{-1}
\end{equation}
We multiply $U$ from the left with an element of the $U(1)$-family of unitaries $(\mathbb{1}+\left(e^{i \alpha}-1\right) \ket{w}\bra{w})$ where $\vert w\rangle$ is an arbitrary fixed normalized vector. Running through all $\alpha \in [0,2\pi)$, we obtain a family of unitaries $U(\alpha)$:
\begin{equation}
    U(\alpha) := U+ \left(e^{i \alpha}-1\right) \ket{w}\bra{w}  U
\end{equation}
By Cayley transforming each of the $U(\alpha)$, we obtain a family of self-adjoint operators $S(\alpha)$
\begin{equation}
    S({\alpha}):=-i \lb U({\alpha}) +\mathbb{1}  \rb \lb U({\alpha})-\mathbb{1}\rb^{-1}
\end{equation}
with $S(0)=S$. We claim that, for any fixed choice of $U$ and $\vert w\rangle$, the Hermitian operators $S(\alpha)$ for varying $\alpha$ differ by a multiple of a rank $1$ projector, i.e., that there exists a normalized vector $\vert v\rangle$ so that for every $\alpha\in [0,2\pi)$ there exists a $\mu(\alpha)\in\mathbb{R}$ obeying:
\be
S(\mu(\alpha)) = S+\mu \vert v \rangle \langle v \vert 
\ee
For the proof, we start with: 
\begin{eqnarray} 
S(\alpha) &=& -i \lb \mathbb{1} + U + \lb e^{i \alpha}-1\rb \ket{w}\bra{w} U\rb  \\ 
 & & \times \lb  U + \lb e^{i \alpha}-1\rb \ket{w}\bra{w} U-\mathbb{1}\rb^{-1}
\end{eqnarray}
Acting from the right with the operator $\lb  U + \lb e^{i \alpha}-1\rb \ket{w}\bra{w} U-\mathbb{1}\rb$ yields:
$$ 
    S(\alpha) \lb  U + \lb e^{i \alpha}-1\rb \ket{w}\bra{w} U-\mathbb{1}\rb 
   $$
\be \mbox{~~~~~~~~~~~~~}=-i \lb 1 + U\rb -i \lb e^{i \alpha}-1\rb \ket{w}\bra{w} U 
\ee
Rearranging the terms:
$$
    S(\alpha) \lb  U -\mathbb{1}\rb + S(\alpha)\lb e^{i \alpha}-1\rb \ket{w}\bra{w} U
$$
\be \mbox{~~~~~~~~~} =-i \lb 1 + U\rb -i \lb e^{i \alpha}-1\rb \ket{w}\bra{w} U 
\ee
After acting with the operator $\lb  U -\mathbb{1}\rb ^{-1}$ from the right we recognize the expression for the Cayley transform $S$ of $U$ and obtain:
\begin{equation}
    S(\alpha)-S = -\lb S(\alpha)+i \rb\lb e^{i \alpha}-1\rb \ket{w}\bra{w} U\lb  U -\mathbb{1}\rb ^{-1} .
\end{equation}
Since both $S(\alpha)$ and $S$ are self-adjoint operators, their difference, $S(\alpha)-S$, is also self-adjoint. This means that the following equation must hold for some $\mu' \lb \alpha \rb \in \mathbb{R}$:
\begin{equation}\label{beta_is_real}
    -\lb S(\alpha)+i \rb\lb e^{i \alpha}-1\rb \ket{w} = \mu'\lb \alpha \rb \left[\lb  U-\mathbb{1}\rb ^{-1}\right]^{\dagger}  U^{\dagger} \ket{w}.
\end{equation}
The left $U(1)$ action on the unitary $U$ which results in $U(\alpha)$ therefore corresponds to the addition of a multiple of a projector $\mu\lb\alpha\rb \ket{v}\bra{v}$ to $S$, resulting in
\begin{equation}
    S(\alpha)=S +\mu\lb\alpha\rb \ket{v}\bra{v},
\end{equation}
where: \be \ket{v}=\frac{1}{\mathcal{N}_v}\lb  U^{\dagger}-\mathbb{1}\rb ^{-1}  U^{\dagger} \ket{w}
\label{where}
\ee
Calculating the norm of \cref{where} while inserting a resolution of the identity, the normalization constant $\mathcal{N}_v$ follows:
\be
\mathcal{N}_v=\sqrt{\sum_m \frac{\abs{w_m}^2}{\abs{u_m-1}^2}}=\frac{\sqrt{\sum_m \abs{w_m}^2 \lb s_m^2+1 \rb}}{2}.
\label{Nv}
\ee
Recall that we use the notation $w_n:=\langle s_n\vert w\rangle, v_n:=\langle s_n\vert v\rangle$ and $U(\alpha)\vert u_n(\alpha)\rangle = u_n(\alpha)\vert u_n(\alpha)\rangle,~~S(\mu)\vert s_n(\mu)\rangle = s_n(\mu)\vert s_n(\mu)\rangle$.

\section{Relating the eigenbases of $U(\alpha)$ and $S(\alpha)$ to those of $U$ and $S$}
\label{finite_sampling_sec}

Our aim is to construct the eigenvectors of $U(\alpha)$ (and therefore of $S(\mu)$) in terms of the eigenbasis of $U$ (and $S$). We have
\begin{equation}
\begin{split}
    u_n(\alpha)\braket{u_n(\alpha)}{u_m}&=\bra{u_n(\alpha)}U(\alpha)\ket{u_m}
    \\
    &=\bra{u_n(\alpha)}(U+(e^{i\alpha}-1)\ket{w}\bra{w} U)\ket{u_m}\\
    &=\braket{u_n(\alpha)}{u_m}u_m\\
    &~~~+(e^{i\alpha}-1)\braket{u_n(\alpha)}{w}\braket{w}{u_m}u_m, \label{versx} 
\end{split}
\end{equation}
and therefore: 
\be
    \braket{u_n(\alpha)}{u_m} =  (e^{i\alpha}-1)\braket{u_n(\alpha)}{w}
     ~\frac{\braket{w}{u_m} u_m}{u_n(\alpha)-u_m} \label{hola}
\ee
Using
\begin{equation}
    1=\braket{u_n(\alpha)}{u_n(\alpha)}=\sum_m\braket{u_n(\alpha)}{u_m}\braket{u_m}{u_n(\alpha)},
\end{equation}
we obtain:
\begin{equation}
    1=\vert e^{-i\alpha}-1\vert^2|\braket{u_n(\alpha)}{w}|^2\sum_m \frac{\abs{u_m}^2  \label{sixteen} \abs{\braket{w}{u_m}}^2}{\abs{u_n(\alpha)-u_m}^2}
\end{equation}
This means that, in \cref{hola}, we can express  $\abs{(e^{i\alpha}-1)\braket{u_n(\alpha)}{w}}$ in terms of known quantities to obtain: 
\begin{equation} \label{rec_ker}
    \abs{\braket{u_n(\alpha)}{u_m}}=\abs{\frac{\braket{w}{u_m}}{u_n(\alpha)-u_m}} \qty(\sum_k \frac{ \abs{\braket{w}{u_k}}^2}{\abs{u_n(\alpha)-u_k}^2})^{-1/2}.
\end{equation}
Via the Cayley transform, this yields: 
\begin{equation}
\abs{\braket{s}{s_n}}=\frac{\abs{v_n}}{\abs{s-s_n}} \lb\sum_k \frac{\abs{v_k}^2} {(s-s_k)^2}\rb^{-1/2}
\end{equation}
Further, the overlap function $\langle s\vert s_n\rangle$ can be chosen real for all $s,s_n$ by suitably choosing the phases of the eigenvectors $\vert s\rangle$. To see this, we calculate:
\be
\frac{s_r(\alpha)-i}{s_r(\alpha)+i}\langle s_n\vert s_r(\alpha)\rangle = \langle s_n\vert U(\alpha)\vert s_r(\alpha)\rangle
\ee
\be
= \langle s_n\vert  \left(  \mathbb{1} + (e^{i\alpha}-1)\vert w\rangle\langle w\vert\right) U \vert s_r(\alpha)\rangle  \nonumber 
\ee
\be = \frac{s_n-i}{s_n+i} \langle s_n\vert s_r(\alpha)\rangle + (e^{i\alpha} - 1)\langle s_n\vert w\rangle\langle w\vert U\vert s_r(\alpha)\rangle   \nonumber 
\ee
This equation can be rewritten as
\be
\langle s_n\vert s_r(\alpha)\rangle = \frac{g_nM(s_r(\alpha))}{s_r(\alpha)-s_n}
\ee
where we separated the $n$-dependent from the $s$-dependent terms by defining: 
\be 
g_n:= (s_n+i)\langle s_n\vert w\rangle
\ee
\be
M(s_r(\alpha)) := \frac{(s_r(\alpha)+i)}{2i}(e^{i\alpha}-1)\langle w\vert U\vert s_r(\alpha)\rangle
\ee
We use the choice of phases of the eigenvectors $\vert s_n\rangle=\vert s_n(\alpha=0)\rangle$ so that $g_n\in \mathbb{R}$ for all $n$. Further, we choose the phases of the eigenvectors $\vert s_r(\alpha)\rangle$ for $\alpha \neq 0$ such that $M(t_r(\alpha))\in \mathbb{R}$ for all $\alpha\neq 0$. Notice that $M(t_r(\alpha))=0$ for $\alpha = 0$. While these choices ensure that $\langle s_n\vert s_r(\alpha)\rangle \in \mathbb{R}$, the overlap function can also be made continuous through a suitable choice of signs, as described in the main text. 

\section{Velocity of the complex eigenvalues}
Differentiating \cref{versx} at $\alpha=0$ for $n=m$, yields:
\be
\left. \frac{du_n(\alpha)}{d\alpha}\right\vert_{\alpha=0} = i~\vert\langle u_n\vert w\rangle\vert^2 u_n
\ee
Since, for any $\alpha$, we could choose $U(\alpha)$ to be the starting unitary $U$ of a new Newton cradle with $\vert v\rangle$, the velocity of the complex eigenvalues for all $\alpha$ reads:
\be
\frac{\dd u_n\lb \alpha \rb}{\dd\alpha}=i \abs{\braket{u_n\lb \alpha\rb} {w}}^2u_n\lb \alpha\rb. \label{compvel}
\ee
From \cref{sixteen}, we have:
\be
\abs{\braket{u_n\lb \alpha\rb} {w}}^2= \frac{1}{\abs{e^{i\alpha}-1}^2} \qty(\sum_m \frac{\abs{w_m}^2}{\abs{u_n(\alpha)-u_m}^2})^{-1}.
\ee
Therefore, \cref{compvel} becomes this differential equation:
\begin{equation}
        \frac{\dd u_n(\alpha)}{\dd \alpha}=i  u_n(\alpha) \frac{1}{\abs{e^{i\alpha}-1}^2} \qty(\sum_m \frac{\abs{w_m}^2}{\abs{u_n(\alpha)-u_m}^2})^{-1}
\end{equation}
\section{Velocity of the real eigenvalues}
The M\"obius transform $u_k=(s_k-i)/(s_k+i)$ yields for the real eigenvalues of $S(\alpha)$:
\begin{equation}
 \frac{d s_n}{d \alpha}= \frac{(s_n+i)^2}{2i} \frac{d u_n}{d \alpha}.
\end{equation}
The inverse M\"obius transform $s_k=i(1+u_k)/(1-u_k)$, applied to \cref{rec_ker}, then yields for the real eigenvalues $\{s_n\}_{n\in 1,...,N}$:
\begin{align}
    \frac{\dd s_n}{\dd \alpha} &=\frac{1}{2} \frac{1}{\abs{e^{i\alpha}-1}^2} \qty(\sum_m \abs{w_m}^2 \frac{s_m^2+1}{4(s_n-s_m)^2})^{-1} \nonumber
    \\
    &=\frac{1}{2} \frac{1}{\sin^2(\frac{\alpha}{2})} \qty(\sum_m \abs{w_m}^2 \frac{s_m^2+1}{(s_n-s_m)^2})^{-1}
\end{align}
\section{Integrating the differential equation for the velocities}
After separating the variables $s_n$ and $\alpha$,
\begin{equation}
    \dd s_n \sum_m \abs{w_m}^2 \frac{s_m^2+1}{(s_n-s_m)^2}=\frac{\dd \alpha}{2\sin^2(\frac{\alpha}{2})}, 
\end{equation}
integrating with the initial conditions $s_n\lb \alpha=0\rb=s_n$ yields:
\begin{equation}
\label{eigs_equation}
    \sum_{m}\abs{w_m}^2 \frac{s_m^2+1}{s_n(\alpha)-s_m}+\sum_{m }\abs{w_m}^2 s_m=\cot(\frac{\alpha}{2})
\end{equation}
As expected, since this yields a polynomial of order $n$ in $s_n(\alpha)$, there are $N$ solutions $\{s_n(\alpha)\}_{n=1,...,N}$.

In order to express \cref{eigs_equation} using the components of $\ket{v}$ rather than $\ket{w}$, we express $\ket{w}$ in terms of $\ket{v}$ and the operator $U$:
\begin{equation}
    \ket{w}=\mathcal{N}_v\  U \lb  U^{\dagger}-\mathbb{1}\rb    \ket{v} = \mathcal{N}_v \lb  \mathbb{1}-U\rb    \ket{v} 
\end{equation}
To obtain the components of $\ket{w}$ and $\ket{v}$, we act with $\bra{u_k}$ from the left:
\begin{equation}
    w_k=\braket{u_k}{w} = \mathcal{N}_v   \bra{u_k}\lb\mathbb{1}-U\rb    \ket{v}, 
\end{equation}
\begin{equation}
    w_k= \mathcal{N}_v \lb  1-u_k\rb v_k 
\end{equation}
Since only the absolute values of $w_k$ appear in \cref{eigs_equation}, we calculate $\abs{w_k}^2$ and then inverse Cayley transform $u_k$ in order to get an expression that depends on the real eigenvalues $\{s_i\}_{i=1,...,N}$:
\begin{eqnarray}
    \abs{w_k}^2&=& \abs{\mathcal{N}_v}^2 \abs{1-u_k}^2 \abs{v_k}^2 \\
    &=& \abs{\mathcal{N}_v}^2 \frac{4}{s_k^2+1} \abs{v_k}^2.
\end{eqnarray}
This yields \cref{eigs_equation} in terms of the $v_k$:
\begin{eqnarray}
\label{eigs_equation_2}
    \cot(\frac{\alpha}{2}) &=&
    \sum_{m} 4 \abs{\mathcal{N}_v}^2\frac{\abs{v_m}^2}{s_n(\alpha)-s_m\lb 0 \rb} \nonumber \\
    & & +\sum_{m }\abs{v_m}^2 4 \abs{\mathcal{N}_v}^2 \frac{s_m}{s_m^2+1} s_m
\end{eqnarray}
\section{The relationship between $\mu$ and $\alpha$}
We claim that $\mu\lb \alpha\rb$ is given by:
\be\label{mu_alpha}
\mu\lb \alpha \rb = \sum_m \abs{w_m}^2 \lb s_m^2+1\rb \lb \cot \lb \frac{\alpha}{2}\rb -\sum_m \abs{w_m}^2 s_m\rb^{-1}
\ee
We start the proof with the observation that 
$S(\alpha)-S = \mu\vert v\rangle\langle v\vert$ yields:
\be
\lb s_n(\alpha)-s_m\rb{\braket{s_m}{s_n\lb\alpha\rb}}
=\mu\lb \alpha\rb {\braket{s_m}{v}\braket{v}{s_n\lb \alpha\rb}}\ee
We solve this equation for $\lb s_n(\alpha)-s_m\rb$ and substitute it into \cref{eigs_equation} to obtain: 
\begin{eqnarray}
\label{mualpha_deriv}
& & \frac{1}{\mu} \sum_m \abs{w_m}^2 \lb s_m^2+1\rb \frac{\braket{s_m}{s_n\lb\alpha\rb}}{\braket{s_m}{v}\braket{v}{s_n\lb \alpha\rb}} \\ & & ~~~~~=\cot\lb\frac{\alpha}{2} \rb-\sum_m \abs{w_m}^2  s_m \label{secnd}
\end{eqnarray}
From \cref{where}, a M\"obius transform yields
\be
\braket{s_m}{v} = \frac{1}{\mathcal{N}_v}\frac{s_m+i}{2i}\braket{s_m}{w},
\ee
i.e.,
\be
\lb s_m+i \rb w_m = 2 i \  \mathcal{N}_v \  v_m.
\ee
Substituting this expression into  Eq.\ref{mualpha_deriv}, we obtain:
\begin{eqnarray}\label{mualpha_deriv2}
& &\frac{1}{\mu} \sum_m 2 i \ \nonumber \mathcal{N}_v  \lb s_m-i \rb w_m^* \frac{\braket{s_m}{s_n\lb\alpha\rb}}{\braket{v}{s_n\lb \alpha\rb}}\\ & & ~~~~~=\cot\lb\frac{\alpha}{2} \rb-\sum_m \abs{w_m}^2  s_m.
\end{eqnarray}
We now insert a resolution of the identity $\mathbb{1}=\sum_m \ket{s_m}\bra{s_m}$ into the inner product $\braket{s_n(\alpha)}{v}$ and use \cref{where} to obtain:
\be
\braket{s_n\lb \alpha\rb}{v}=\frac{1}{\mathcal{N}_v} \sum_m \braket{s_n\lb \alpha\rb}{s_m} \braket{s_m}{w}\frac{u_m^*}{u_m^*-1}
\ee
We then divide by $\braket{s_n(\alpha)}{v}$ and complex conjugate the entire expression:
\be
1=\frac{1}{\mathcal{N}_v}\sum_m w_m^*\frac{s_m-i}{-2i} \frac{\braket{s_m}{s_n\lb\alpha\rb}}{\braket{v}{s_n\lb \alpha\rb}}
\ee
Now we recognize the same sum as in  Eq. \ref{mualpha_deriv2}, up to constants, so we can finally express $\mu$ as a function of $\alpha$:
\be
\mu=4 \mathcal{N}_v^2 \lb \cot\lb\frac{\alpha}{2} \rb-\sum_m \abs{w_m}^2 s_m\rb^{-1} \label{mua}
\ee
With ${\mathcal{N}_v}$ given in \cref{Nv}, this yields \cref{mu_alpha} as claimed.

\section{Reformulation of the characteristic equation of $S(\mu)$}
We now derive the equation:
\begin{equation}
    \sum_{m}\frac{\abs{v_m}^2} {s_n(\mu)-s_m}=\frac{1}{\mu}. \label{twotwo}
\end{equation}
To this end, we use
$
    \abs{w_k}^2= \abs{\mathcal{N}_v}^2 \frac{4}{s_k^2+1} \abs{v_k}^2
$
to express $w_m$ in terms of $v_m$ in \cref{eigs_equation}:
\begin{equation}
\label{intermediate_selfadj}
    4 \abs{\mathcal{N}_v}^2 \sum_{m} \frac{\abs{v_m}^2}{s_n(\alpha)-s_m} =\cot(\frac{\alpha}{2})-\sum_{m }\abs{w_m}^2 s_m,
\end{equation}
Using \cref{mua}, we express the right-hand side in  Eq.\ref{intermediate_selfadj} in terms of $\mu(\alpha)$:
\be
 4 \abs{\mathcal{N}_v}^2 \sum_{m} \frac{\abs{v_m}^2}{s_n(\alpha)-s_m}= 4 \abs{\mathcal{N}_v}^2 \frac{1}{\mu(\alpha)}.
\ee
Finally, dividing by $4 \abs{\mathcal{N}_v}^2$, and expressing the dependence on $\alpha$ as a dependence on $\mu$, we obtain Eq.\ref{twotwo}. We remark that an equation equivalent to  Eq.\ref{twotwo}, which is equivalent to the characteristic equation $\det(S(\mu)-s\mathbb{1})=0$ of $S(\mu)$, was used in [49] 
to derive a numerical algorithm for the eigenvalue problem, along with a perturbative stability analysis. 
\bigskip\newline
\section{Calculating the eigenvalues $s_n(\mu)$ as a function of $\mu$ by using the Lagrange inversion theorem to invert $\mu(s)$}
From \cref{twotwo}, we obtain for every eigenvalue $s$ the corresponding value $\mu$:
\begin{equation}
\mu(s) = \left(\sum_{m=1}^N\frac{\abs{v_m}^2} {s-s_m}\right)^{-1}
\end{equation}
Vice versa, given a value for $\mu$, we now calculate the eigenvalues $s_r(\mu)$ nonperturbatively in the sense that we obtain $s_r(\mu)$ as a Taylor series in $\mu$ about $\mu=0$ to all orders:
\begin{equation}
    s_r(\mu) = s_r + \sum_{n=1}^\infty G_n \frac{\mu^n}{n!}
\end{equation}
We use the Lagrange inversion theorem in order to calculate explicit expressions for $G_n ~ \forall n\ge 1$. To this end, we need to evaluate:
\begin{eqnarray}
    G_n &=& \lim_{s\rightarrow s_r} \frac{d^{n-1}}{ds^{n-1}}\left(\frac{s-s_r}{\mu(s)}\right)^n \\
    &=& \lim_{s\rightarrow s_r} \frac{d^{n-1}}{ds^{n-1}}\left(\sum_{n=1}^N \vert v_n\vert^2~\frac{s-s_r}{s-s_n}\right)^n 
\end{eqnarray}
Defining
\begin{equation}
    R(s):= \sum_{n=1}^N\vert v_n\vert^2~\frac{s-s_r}{s-s_n}
\end{equation}
we have:
\begin{equation}
    \lim_{s\rightarrow s_r} R(s) = \vert v_r\vert ^2 \label{g0}
\end{equation} 
After a short calculation, we obtain for the $m$'th derivative $R^{(m)}(s)$ of $R(s)$ for all $m \ge 1$: 
\begin{equation}
    \lim_{s\rightarrow s_r} R^{(m)}(s) = \sum_{\substack{ n=1 \\(n\neq r)}}^N~\frac{(-1)^{m+1}m!\vert v_n\vert^2}{(s_r-s_n)^m} \label{g1}
\end{equation}
Since $\sum_{n=1}^N\vert v_n\vert^2 = 1$ we can combine \cref{g0,g1} to obtain one equation for all $m\ge 0$:
\begin{equation}
    \lim_{s\rightarrow s_r} R^{(m)}(s) = \sum_{\substack{ n=1 \\(n\neq r)}}^N~\frac{(-1)^{m+1}m!\left(\vert v_n\vert^2-\delta_{m,0}/(N-1)\right)}{(s_r-s_n)^m}
\end{equation}
Using the general Leibniz rule, we have:
\begin{eqnarray}
    G_n &=& \lim_{s\rightarrow s_r} \frac{d^{n-1}}{ds^{n-1}}R(s)^n \\
    &=& \lim_{s\rightarrow s_r}\sum_{\substack{k_1,...,k_n=0\\ \left(\sum_{i=1}^n k_i=n-1\right)}}^{n-1}\frac{(n-1)!R^{(k_1)}\cdot...\cdot R^{(k_n)}}{k_1!\cdot ...\cdot k_n!}\nonumber
\end{eqnarray}
A straightforward calculation then yields the $G_n$ explicitly for all $n\ge 0$:
\begin{equation}
    G_n = -(n-1)!\sum_{\substack{k_1,...,k_n=0\\ \left(\sum_{i=1}^n k_i=n-1\right)}}^{n-1}\sum_{\substack{p_1,...,p_n=1\\ (p_1\neq r,...,p_n\neq r)}}^N\prod_{i=1}^n\frac{\vert v_{p_i}\vert^2-\frac{\delta_{k_i,0}}{N-1}}{(s_r-s_{p_i})^{k_i}} 
 \end{equation}
We finally obtain:
\begin{equation}
s_r(\mu) = s_r -\sum_{n=1}^\infty \frac{\mu^n}{n}\!\!\!\!\!\!
\sum_{\substack{k_1,...,k_n=0\\ \left(\sum_{i=1}^n k_i=n-1\right)}}^{n-1}\sum_{\substack{p_1,...,p_n=1\\ (p_1\neq r,...,p_n\neq r)}}^N\prod_{i=1}^n\frac{\vert v_{p_i}\vert^2-\frac{\delta_{k_i,0}}{N-1}}{(s_r-s_{p_i})^{k_i}} 
 \end{equation}
 
For example:
\begin{equation}
    s_r(\mu) = s_r + \vert v_r\vert^2 \mu - \vert v_r\vert ^2 \left(\frac{\vert v_r\vert^2}{2}-\sum_{p\neq r}\frac{\vert v_p\vert^2}{s_r-s_p}\right)~\mu^2 +...
\end{equation}
\bigskip\newline
\section{Cauchy interlacing} The Cauchy interlacing theorem arises as a special case. The theorem states that the $N-1$ eigenvalues $a_1<a_2<\ldots <a_{N-1}$ of any Hermitian $(N-1)\times (N-1)$ matrix $A$ obtained by deleting the $r$-th row ($r$ is arbitrary) and $r$-th column of a Hermitian $N\times N$ matrix $S$ with nondegenerate spectrum are interlaced in the $N$ eigenvalues of $S$, i.e.,: $s_n<a_n<s_{n+1}~\forall n=1\ldots (N-1)$. It is straightforward to show that one obtains this result as the special case of the Newton cradle of $S$ in which $\vert v\rangle$ is chosen to be the vector with the components $\delta_{r,i}$ in the basis in which the matrix $S$ is given, and letting $\mu\rightarrow \infty$. Using Eq.2 of the main text for $\mu\rightarrow \infty$, we can conclude more, namely that each of the interlaced eigenvalues $a_n$ is a solution of Eq.3 of the main text, which means that we can identify the interlaced eigenvalues as $a_n=s_n^*$ for $n=1,...~,N-1$. 

\section{Recovery of ``Eigenvectors from eigenvalues" result by Denton et al}

Finally, we show that the well-known \it Eigenvectors from Eigenvalues \rm result by Denton et al, \cite{Denton_2021}, follows from (2). In (2), as mentioned in the section on the Cauchy interlacing theorem, taking the limit $\mu \rightarrow \infty$, we effectively delete a row and a column in a basis that contains the vector $\ket{v}$ (1). We then obtain an expression for the components of the eigenvectors $\{\ket{s_n}\}_n$ of $S$ in a basis vector $\ket{v}$:

\begin{equation}
\sum_{m=1}^N\frac{\abs{\bra{v}\ket{s_m}}^2} {s^*-s_m}=0. 
\end{equation}

Viewing $\abs{\bra{v}\ket{s_m}}^2$ as a complex function of $s^*$, and noting that the right hand side has no poles, we conclude that it must be proportional to this product:

\begin{equation}
    \abs{\bra{v}\ket{s_m}}^2 \propto \prod_{i=1}^{n-1} \lb s_i^*-s_m\rb.
\end{equation}

Finally, requiring $\ket{s_m}$ to be normalized vectors and that when $\ket{v}\rightarrow \ket{s_m} \Rightarrow s^*\rightarrow s$, we  obtain:

\begin{equation}
     \abs{\bra{v}\ket{s_m}}^2 = \frac{\prod_{i=1}^{n-1} \lb s_i^*-s_m\rb}{\prod_{k=1,k\neq m}^{n-1} \lb s_k-s_m\rb}.
\end{equation}

This reproduces the corresponding expression in \cite{Denton_2021}.


\end{document}